\documentclass[11pt,letterpaper]{llncs}
\usepackage{amsmath}
\usepackage{amssymb}
\usepackage{graphicx}
\usepackage{marvosym}
\usepackage{url}
\usepackage{fancyhdr}
\usepackage{graphicx}

\usepackage{geometry}
\newcommand{\keywords}[1]{\par\addvspace\baselineskip
\noindent\keywordname\enspace\ignorespaces#1}

\fancypagestyle{firstpage}{\fancyhf{}
\setcounter{page}{33}
\rfoot{\thepage}
}
\begin{document}


\title{\LARGE{Attention-Based Efficient Breath Sound Removal in Studio Audio Recordings}}








%
%
\author{\large{Nidula Elgiriyewithana\textsuperscript{*} \and N D Kodikara\textsuperscript{*}}}

\institute{\large{ Robert Gordon University, Aberdeen, Scotland \and Informatics Institute of Technology, Colombo, Sri Lanka}}

%


%
%


\maketitle

\thispagestyle{firstpage}

\begin{abstract}
In this research, we present an innovative, parameter-efficient model that utilizes the attention U-Net architecture for the automatic detection and eradication of non-speech vocal sounds, specifically breath sounds, in vocal recordings. This task is of paramount importance in the field of sound engineering, despite being relatively under-explored. The conventional manual process for detecting and eliminating these sounds requires significant expertise and is extremely time-intensive. Existing automated detection and removal methods often fall short in terms of efficiency and precision. 
Our proposed model addresses these limitations by offering a streamlined process and superior accuracy, achieved through the application of advanced deep learning techniques. A unique dataset, derived from Device and Produced Speech (DAPS), was employed for this purpose. The training phase of the model emphasizes a log spectrogram and integrates an early stopping mechanism to prevent overfitting. 
Our model not only conserves precious time for sound engineers but also enhances the quality and consistency of audio production. This constitutes a significant breakthrough, as evidenced by its comparative efficiency, necessitating only 1.9M parameters and a training duration of 3.2 hours - markedly less than the top-performing models in this domain. The model is capable of generating  identical outputs as previous models with drastically improved precision, making it an optimal choice.

\end{abstract}


\section{Introduction}\label{sec:Introduction}

Sound engineering encompasses a wide array of activities, including the manipulation and production of audio signals for diverse applications such as music, speech, film, and broadcasting. A prevalent challenge in this field is the occurrence of non-speech vocal sounds in audio recordings, such as breaths and lip smacks. These sounds can be distracting and may degrade the quality and intelligibility of the audio\cite{owsinski2014mixing}, particularly when the recording is intended for professional or academic use.Historically, this issue has been addressed through manual editing of audio waveforms or the use of noise gates, which dynamically reduce signals below a certain threshold by adjusting gain and ratio settings\cite{terrell2011automatic}. However, these methods pose their own challenges, including substantial time consumption and the need for domain expert knowledge\cite{amatriain2011dafx}.

\begin{flushleft}
\footnotesize{David C. Wyld et al. (Eds): DMML, CSITEC, NLPI, BDBS - 2024 \\
pp. 49-58, 2024. CS \& IT - CSCP 2024}
\hfill
\footnotesize{\text{DOI: 10.5121/csit.2024.140604}}
\end{flushleft}

The manual removal of non-speech vocal sounds is a meticulous and time-consuming task that demands significant expertise and focus from the sound engineer. It involves a detailed examination of the audio waveform, identification of undesired sounds, and their elimination using specialized software tools\cite{owsinski2014mixing}. This method can inadvertently introduce artifacts or distortions into the audio, thereby affecting its authenticity and quality. Furthermore, manual removal is impractical for tasks requiring rapid and efficient audio processing, such as managing lengthy or multiple recordings simultaneously\cite{ramirez2021differentiable}.
 The automatic removal of non-speech vocal sounds presents a more efficient and desirable alternative, potentially enhancing the quality and consistency of audio while conserving the time and effort of sound engineers. Despite its importance, research on the automatic removal of non-speech vocal sounds is surprisingly limited. Most of the existing studies have focused on related but distinct problems, such as noise reduction, speech enhancement, or vocal separation\cite{mehrish2023review}. Only a few studies have explicitly addressed the problem of non-speech vocal sound removal.
 
While previous methodologies have shown promise, there still exists a need for a more efficient and effective model. In this research, we introduce an innovative, parameter-efficient model that utilizes the attention U-Net architecture for the automatic detection and eradication of non-speech vocal sounds, specifically breath sounds, in vocal recordings. This task is of paramount importance in the field of sound engineering, and our model addresses the limitations of existing methods by offering a streamlined process and superior accuracy, achieved through the application of advanced deep learning techniques.

We begin by providing a comprehensive background on sound engineering and the necessity for effective removal of non-speech vocal sounds in audio recordings ~\ref{sec:Introduction}.Reviewing existing methods, we elaborate on their shortcomings and detail our approach which employs a unique dataset and cutting-edge deep learning methods ~\ref{sec:Relatedwork} ~\ref{sec:Methodology}. We then reveal the results from our innovatory approach, displaying its potential for saving valuable time for sound engineers while boosting the quality and consistency of audio production ~\ref{sec:Results}. Further, we explore its real-time application and the possible influence it could wield on sound engineering, underscoring its superior accuracy compared to earlier models ~\ref{sec:Discussion}. Ultimately, we ponder over the impacts and advancements our research has contributed to the field of sound engineering in the conclusion ~\ref{sec:Conclusion}. Our model symbolizes a momentous stride towards the automatic detection and extermination of non-speech vocal sounds, presenting an efficient and robust resolution to a perennial challenge.

\section{Related work}\label{sec:Relatedwork}

In recent years, there has been growing interest in developing algorithms for automatic detection and removal of breath sounds from speech signals, and several studies have explored automated methods for detecting and removing breath sounds from audio recordings.In 2007 Ruinskiy and Lavner presented an algorithm aimed at accurately detecting breaths in speech or song signals using template matching based on mel frequency cepstral coefficients (MFCCs) \cite{ruinskiy2007effective}. Their approach achieved a high correct identification rate of 98\% with a specificity of 96\%.
 
In 2009, Rapcan and Reilly developed an algorithm for both detecting and removing breath sounds from speech signals, particularly focusing on its impact on cognitive studies of speech and language\cite{rapcan2009use}. Their approach resulted in a notable increase in discrimination ability, demonstrating the potential of automated breath removal in enhancing classification accuracy.
 
Magdalena and Zi6lko proposed an algorithm based on wavelet decomposition for automatic detection of breath events in speech signals in 2013\cite{igras2013wavelet}. Their method incorporated temporal features and dynamic time warping to achieve robust breath detection, with applications extending to speech recognition systems.
 
Dumpala and Alluri 2017 introduced an algorithm for automatic detection of breath sounds in spontaneous speech, emphasizing its significance in speaker recognition systems\cite{dumpala2017algorithm}. Their rule-based approach outperformed previous methods, highlighting the importance of mitigating the impact of breath sounds on speaker recognition accuracy.
 
Detecting breathing sounds in realistic Japanese telephone conversations was explored by Takashi et al. in 2018\cite{fukuda2018detecting}. They proposed a method that leverages acoustic information specialized for breathing sounds, leading to a two-step approach that can detect breath events with an accuracy of 97.4\%.
 
Recently, there has been growing interest in using deep learning techniques for speech processing tasks. Marco et al. proposed a differentiable signal processing framework for black-box audio effects in 2021\cite{ramirez2021differentiable}. Their approach trains a model with a multi-band noise gate Fx layer to automatically remove breath sounds and other non-speech vocalisations from speech signals.
 
While significant progress has been made in breath sound detection and removal, there remains a need for more efficient and advanced methods that can handle complex speech signals and provide better accuracy. Our proposed method aims to address this challenge by introducing a novel deep learning architecture that can effectively detect and remove breath sounds from speech signals without damaging the original recording.

\section{Methodology}\label{sec:Methodology}

Our study proposes a comprehensive methodology that includes a meticulously designed model, a robust model training regimen, and a specialized U-Net model. Each of these components is crucial in addressing the complex task of eliminating breath sounds from vocal recordings.

\subsection{Model Architecture }

The backbone of our study is the specialized U-Net model emphasized with attention mechanism\cite{oktay2018attention}. Originally envisioned for biomedical image segmentation, the U-Net model has been noted for its capacity to capture both local features and expansive context within an image. In our case, this image is the spectrogram, derived from the Short Time Fourier Transform (STFT) of the audio files.  
The Short Time Fourier Transform (STFT) can be represented as:
\begin{equation}
STFT(x) = \sum_{n=-\infty}^{\infty} x[n] \cdot w[n - m] \cdot e^{-j\omega m}
\end{equation}

where \(x[n]\) is the input signal, \(w[n - m]\) is the window function, and \(m\) is the time index.The STFT renders the audio file as a 2D array, segregating and capturing nuanced patterns along temporal and frequency domains\cite{griffin1984signal}.Although the STFT yields a complex signal, our methodology focuses solely on its amplitude. The amplitude undergoes processing through a softmask followed by traversal through our 
tailored U-Net algorithm\cite{shimauchi2017relationships}.
\begin{figure}[hbt!]
\centering
\includegraphics[width=0.95\textwidth]{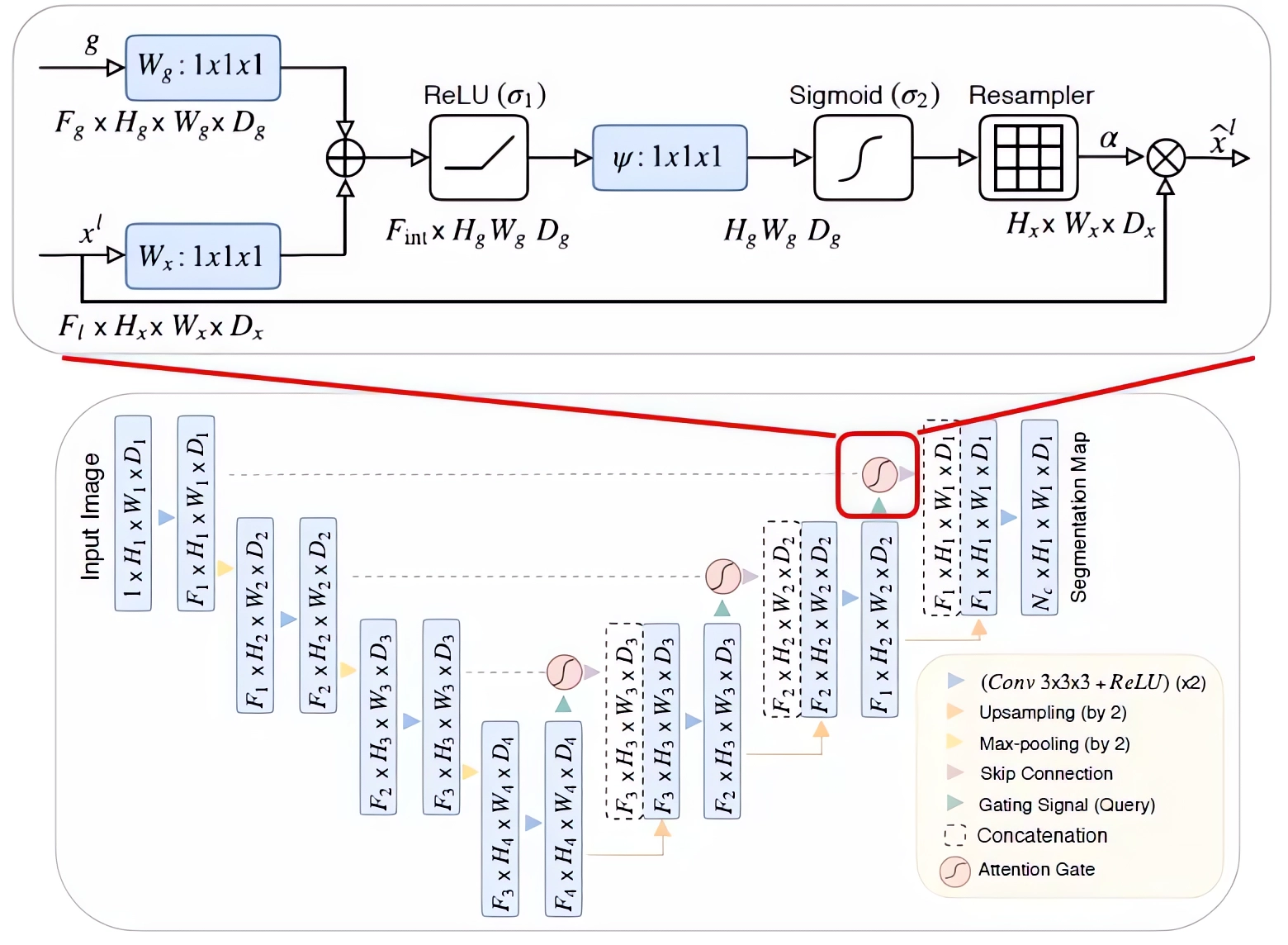}
\caption{Model Architecture}
\label{fig:Model_Architecture}
\end{figure}

\vspace{2cm}

The U-Net model possesses an input shape of 3872x2048x1, uses batch normalization, has a singular class, uses 16 filters, includes 4 layers, incorporates a dropout rate of 20 percent, and employs a sigmoid function for output activation. These parameters were meticulously chosen to obtain the desired results.

\subsection{Loss Function}

The design of the loss function is pivotal in the training of the network model. The loss function essentially quantifies the discrepancy between the predicted output and the actual output. The objective of the training process is to minimize this discrepancy. In this research, we have employed a custom loss function for the signal enhancement task. This loss function is unique in that it not only incorporates a Mean Absolute Error (MAE) term but also an additional term that specifically focuses on preserving speech over reducing noise. The Mean Absolute Error (MAE) is a popular metric in regression problems and is defined as the average of the absolute differences between the predicted and actual values\cite{toro1968test}. In our context, the MAE term is given by:

\begin{equation}
 MAE = \frac{1}{N} \sum_{i=1}^{N} |y_{true,i} - y_{pred,i}| 
\end{equation}

Where \(y_{true,i}\) and \(y_{pred,i}\) are the true and predicted amplitude values of the speech signal, respectively, and N is the total number of samples.

The second term in our loss function is designed to prioritize the preservation of speech over noise reduction. This is achieved by penalizing the model more heavily when it fails to correctly predict the amplitude of the speech signal. Mathematically, this is expressed as:

\begin{equation}
 SpeechLoss = 2 * \frac{1}{N} \sum_{i=1}^{N} |(y_{true,i})^2 - y_{pred,i} * y_{true,i}| 
\end{equation}

The final loss function is a sum of the MAE and the speech loss terms:

\begin{equation}
 Loss = MAE + SpeechLoss 
\end{equation}

The use of this custom loss function ensures that the model not only minimizes the overall error in amplitude prediction but also prioritizes the preservation of speech components over noise reduction.

\subsection{Dataset}

The underlying data for our methodology comes from the Device and Produced Speech (DAPS) Dataset \cite{6981922}. The DAPS dataset houses 100 authentic and clean speech recordings, scrupulously cleaned of breaths and lip smacks. For effective learning and unbiased evaluation, the data is partitioned into training, validation, and testing repositories. Sizes for these sets are 213.5, 30.2, and 23.8 minutes respectively.

\subsection{Model Training }

The model training phase revolves around a log spectrogram which operates on a 22,050Hz sampling rate. Our approach necessitates a frame length of 4096 and a frame step of 512. Additional parameters include a 'hann' window function and the setting of 'center' to True for more precise frequency representation. To facilitate a quick and precise learning process, an V100 GPU with its high computation power is exploited. The training procures the use of Adam optimization and incorporates an early stopping mechanism that continuously scrutinizes validation loss to ensure the model's prevention from overfitting and assures achievement of an optimal state.

\section{Results}\label{sec:Results}

In this results section, we detail the performance of our model that’s designed to remove non-speech sounds, such as breaths in vocal recordings. We compare it with two of the best models currently available, Inception and MobileNetV2, referenced in the work of Ramirez et al. \cite{ramirez2021differentiable}. Our model’s accuracy is measured by its ability to identify and remove breath sounds from the recordings, which is presented as a percentage. Along with accuracy, we also use the Mel-Frequency Cepstral Coefficients (MFCC) Distance to measure spectral precision, and the Perceptual Evaluation of Speech Quality (PESQ) to assess the clarity of speech. The detailed results are shown in Table \ref{tab1}, highlighting our model’s effectiveness in improving the quality of vocal recordings by removing unwanted breath sounds.

\begin{table}[htbp]
\begin{center}
\begin{tabular}{|p{4cm}|p{4cm}|p{2cm}|p{2cm}|}
\hline
\textbf{Model} & \textbf{MFCC Distance} & \textbf{PESQ} & \textbf{Accuracy} \\
\hline
Proposed Model & 0.0371 & 3.8433  & 97\%\\
\hline
Inception & 0.0186 & 3.9452 &  98\% \\
\hline
MobileNetV2& 0.0231 & 3.9448 &  97\% \\
\hline
\end{tabular}
\end{center}
\caption{Performance comparison on various metrics}
\label{tab1}
\end{table}

The MFCC Distance for the proposed model is 0.0371, which is slightly higher than the Inception and MobileNetV2 models, which have an MFCC Distance of 0.0186 and 0.0231 respectively. The MFCC Distance is a measure of the dissimilarity between the original and processed signals, with a lower value indicating a closer match to the original signal. Therefore, while our proposed model does not outperform the Inception and MobileNetV2 models in terms of MFCC Distance, the difference is not substantial, indicating that our model is still highly competitive.The Perceptual Evaluation of Speech Quality (PESQ) scores for our proposed model and the other two models are quite close, with our model having a PESQ score of 3.8433, slightly lower than the Inception's score of 3.9452 and MobileNetV2's score of 3.9448. PESQ is a metric that assesses the perceived speech quality after the sound processing, with a higher score indicating better perceived quality.  Most importantly, the accuracy of our proposed model in removing breath sounds from vocal recordings is commendable, with a score of 97\%. This is on par with the MobileNetV2 model and only slightly less than the Inception model, which has an accuracy of 98\%.

\begin{figure}[hbt!]
\centering
\includegraphics[width=\textwidth]{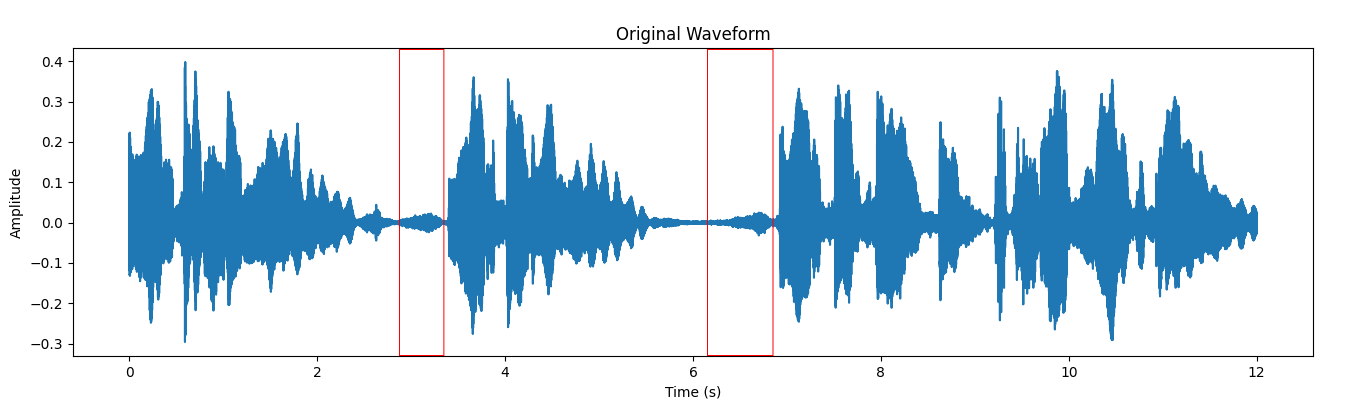}
\caption{Original Waveform}
\label{fig:original_waveform}
\end{figure}

\begin{figure}[htbp]
\centering
\includegraphics[width=\textwidth]{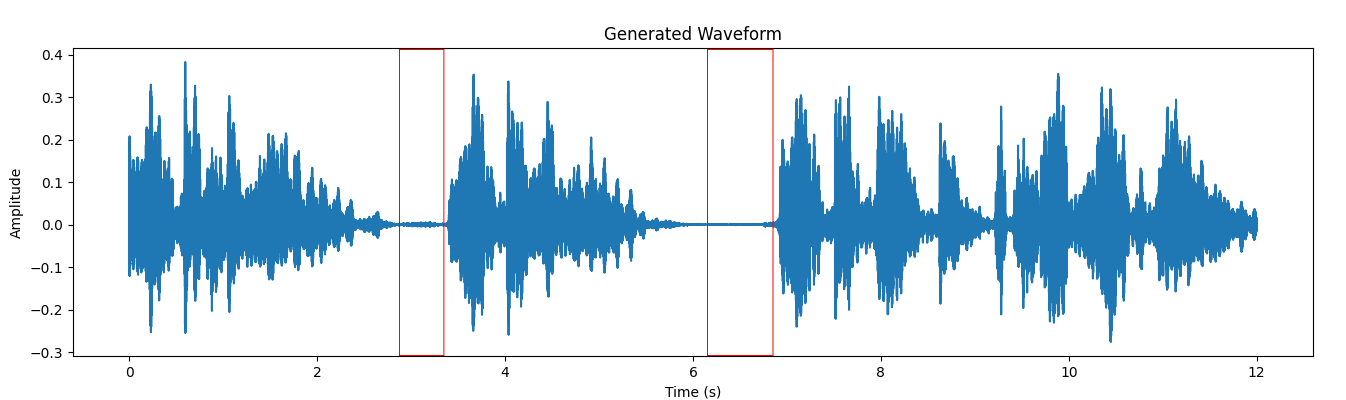}
\caption{Waveform after Breath Removal}
\label{fig:waveform_after_removal}
\end{figure}

Figures \ref{fig:original_waveform} and \ref{fig:waveform_after_removal} demonstrate the original waveform and the waveform post the removal of the breath from the vocal recording. The highlighted box in these figures signifies the existence of breath sound.

\begin{figure}[htbp]
\centering
\includegraphics[width=1\textwidth]{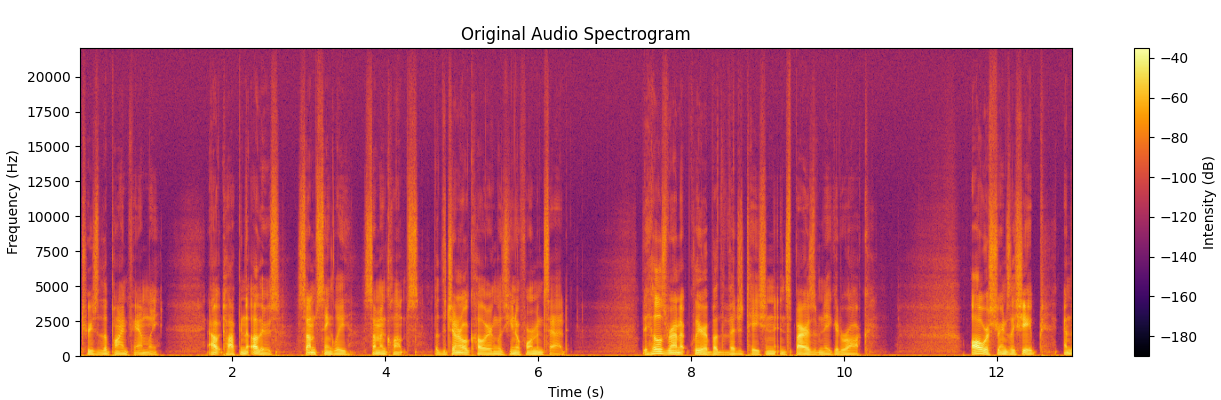}
\caption{Original Spectrogram}
\label{fig:original_spectrogram}
\end{figure}

\begin{figure}[htbp]
\centering
\includegraphics[width=1\textwidth]{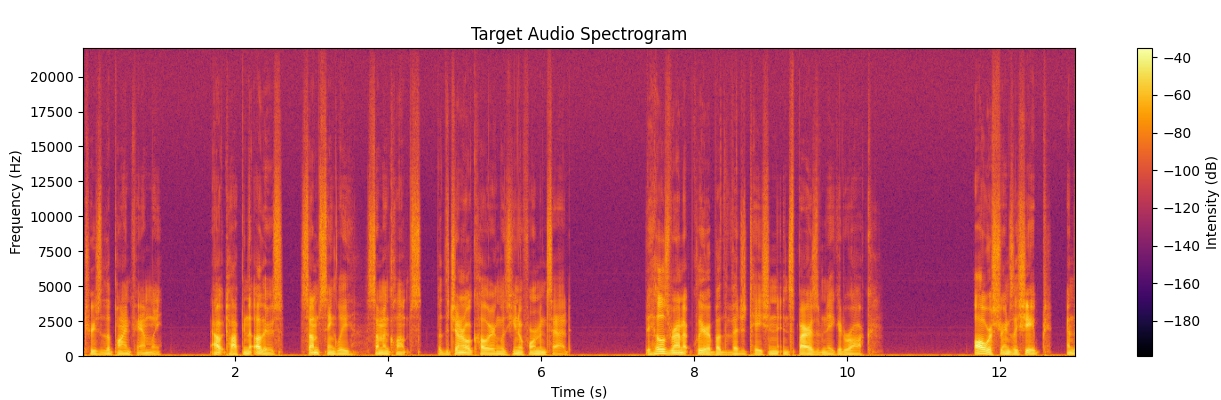}
\caption{Target Spectrogram}
\label{fig:target_spectrogram}
\end{figure}

\begin{figure}[htbp]
\centering
\includegraphics[width=1\textwidth]{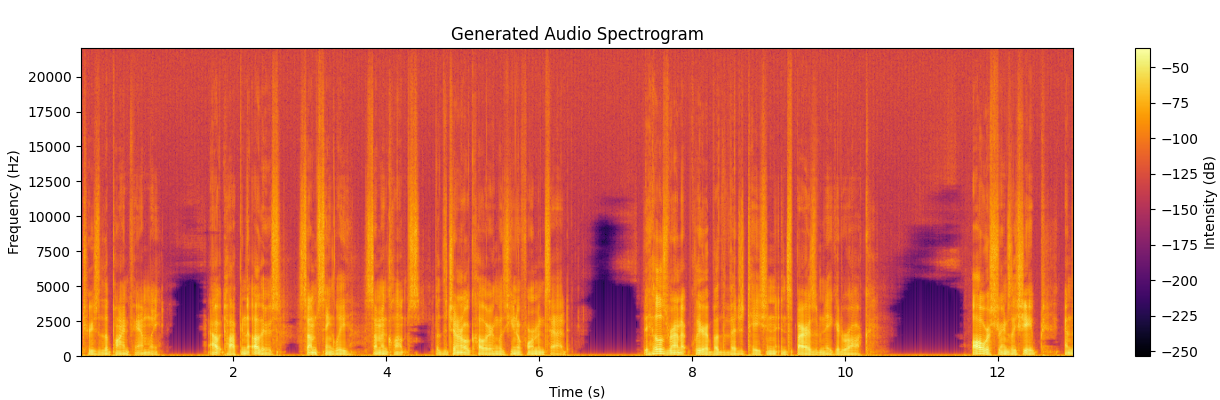}
\caption{Model Generated Spectrogram}
\label{fig:model_spectrogram}
\end{figure}

Figures 4, 5, and 6 represent the spectrogram view of the original audio, the
 target audio (manually breath sound removed vocal recording), and the model
generated vocal recording, respectively. This demonstrates the model’s ability to
filter out unwanted noise without affecting the overall quality of the recording.

Table \ref{tab2} compares the number of parameters, epochs, and time taken for each model. 

\begin{table}[htbp]
\begin{center}
\begin{tabular}{|p{5cm}|p{2cm}|p{2cm}|p{2cm}|}
\hline
\textbf{Model} & \textbf{Parameters} & \textbf{Epochs} & \textbf{Time (hours)}\\
\hline
Proposed Model & 1.9M & 60 & 3.2\\
\hline
Inception & 2.8M & 89 & 7.4\\
\hline
MobileNetV2 & 2.2M & 60 & 4.8\\
\hline
\end{tabular}
\end{center}
\caption{Comparison of the number of epochs and time taken}
\label{tab2}
\end{table}

Our proposed model is efficient, with fewer parameters (1.9M) and less time taken (3.2 hours) for training compared to the other models\cite{ramirez2021differentiable}
. This underlines the efficiency of our approach, which is able to generate nearly identical output to previous models, but with a more streamlined process.

\section{Discussion and Future Work}\label{sec:Discussion}
The study showcases a parameter-efficient model that utilizes deep learning to accurately detect and remove non-speech vocal sounds like breath noises from vocal recordings. The model excels in post-processing, reducing manual work significantly thus, benefitting sound engineering. However, the real potential lies in extending the model to real-time systems such as live broadcasts or interactive voice response systems. In spite of using fewer parameters, the model yielded better results than existing models, as confirmed by audio quality metrics. Future work could focus on devising objective methods to evaluate sound removal, exploring other deep learning approaches, and expanding the dataset to include varied vocal recordings for broader non-speech sound management.

\section{Conclusion}\label{sec:Conclusion}
This research has introduced a parameter-efficient model, which utilizes the attention U-Net architecture for the automatic detection and removal of breath sounds in vocal recordings. Demonstrating a significant need for research in this area in the sound engineering field, the paper emphasizes the inefficiencies and inaccuracies of previous automated methods. Our approach, characterized by its innovation and superior performance, leverages advanced deep learning techniques and a carefully constructed methodology for effectively capture local and expansive context within the spectrogram derived from the STFT of the audio file, thereby enabling accurate and efficient detection and removal of breath sounds. Therefore, our research holds significant pertinence in the sound engineering domain and can be used by voice-over artists, singing artists, and other audio recording professionals to improve the quality of their audio files efficiently without the inconvenience of manual editing.



\bibliographystyle{ieeetr}
\bibliography{Reference}

\begin{flushleft}
\footnotesize{\textcopyright{} 2024 By {AIRCC Publishing Corporation}. This article is published under the Creative Commons Attribution (CC BY) license.}
\end{flushleft}



\end{document}